\documentclass[reprint,aps,prd,superscriptaddress,preprintnumbers,nofootinbib]{revtex4-1}

\usepackage{physics}
\usepackage{amsfonts}
\usepackage{amsmath}
\usepackage{gensymb}
\usepackage{amsthm}
\usepackage{amssymb}
\usepackage{array}
\usepackage{dcolumn}
\usepackage[svgnames]{xcolor}
\usepackage[hidelinks]{hyperref}
\hypersetup{
    colorlinks=true,
    linkcolor=NavyBlue,
    filecolor=NavyBlue,
    urlcolor=NavyBlue,
    citecolor=NavyBlue,
    }

\usepackage{graphics}
\usepackage{graphicx}
\usepackage{adjustbox}
\usepackage{breqn}
\usepackage{float}

\def\be{\begin{equation}}
\def\ee{\end{equation}}
\def\bea{\begin{eqnarray}}
\def\eea{\end{eqnarray}}

\newcommand{\kpc}{\rm kpc}

\makeatletter
\let\cat@comma@active\@empty
\makeatother

\begin{document}

\title{Discriminating Dark Matter Origins with Directional Detection}

\author{Nicole F.~Bell}
\affiliation{ARC Centre of Excellence for Dark Matter Particle Physics, \\
School of Physics, The University of Melbourne, Victoria 3010, Australia}

\author{Chiara Lisotti}
\email{maria.lisotti@sydney.edu.au}
\affiliation{ARC Centre of Excellence for Dark Matter Particle Physics, \\
School of Physics, The University of Sydney, NSW 2006, Australia}

\author{Jayden L.~Newstead}
\affiliation{ARC Centre of Excellence for Dark Matter Particle Physics, \\
School of Physics, The University of Melbourne, Victoria 3010, Australia}

\author{Ciaran A.~J.~O'Hare}
\affiliation{ARC Centre of Excellence for Dark Matter Particle Physics, \\
School of Physics, The University of Sydney, NSW 2006, Australia}

\author{Iman Shaukat Ali}
\email{ishaukatali@student.unimelb.edu.au}
\affiliation{ARC Centre of Excellence for Dark Matter Particle Physics, \\
School of Physics, The University of Melbourne, Victoria 3010, Australia}

\begin{abstract}
Scenarios where dark matter is boosted to relativistic velocities provide a promising probe of sub-GeV dark matter. Cosmic-ray upscattered and supernova-produced dark matter generate relativistic fluxes peaked toward the Galactic Centre, an anisotropy that offers a strong directional signature and is not mimicked by any terrestrial or cosmic background. We determine how many directional recoil events are required in a gas time-projection chamber to distinguish various scenarios for the origin of dark matter particles arriving in the solar system, which are otherwise indistinguishable without directionality. We find that standard halo dark matter particles can be distinguished from boosted populations with as few as $\mathcal{O}(20)$ events under reasonable track reconstruction performance and background conditions. 
%
\end{abstract}

\maketitle  

\section{Introduction}

The existence of dark matter (DM) is established through its gravitational effects at galactic (and larger) scales, but the properties of its particle nature remain unknown. Direct detection experiments search for nuclear or electronic scattering signatures, and have placed stringent constraints on the parameter space of Weakly Interacting Massive Particle (WIMP)-like dark matter, in the $\sim$GeV-TeV mass range {\cite{LZ:2024zvo,XENON:2024hup,CRESST:2019jnq,XENON:2026qow}}. However, even detectors with low-energy thresholds lose sensitivity to sub-GeV DM, due to insufficient energy deposition in the scattering process. The strongest constraints on sub-GeV dark matter arise instead from cosmology and astrophysical observations, and are some $\sim$15 orders of magnitude weaker than those from direct detection of GeV-mass WIMPs~\cite{Nadler:2019zrb,Boddy:2018kfv,Gluscevic:2017ywp,Slatyer:2018aqg,Xu:2018efh,Bhoonah:2018wmw,Bhoonah:2018gjb,Wadekar:2019mpc}. 

The sensitivity of experiments to low-mass DM can be enhanced by considering mechanisms that may boost the energy of particles arriving at the detector, for example if the DM interacts with and reflects off the solar interior \cite{An:2017ojc,An:2021qdl,Emken:2017hnp,Chen:2020gcl,Emken:2021lgc,Kouvaris:2015nsa, PandaX:2024syk}, or scatters off energetic particles from blazar jets \cite{Granelli:2022ysi,Wang:2021jic,Bhowmick:2022zkj, Jeesun:2025gzt, Wang:2025ztb, DeMarchi:2025uoo, CDEX:2024qzq, Xia:2024ryt,Barillier:2025xct}. A flux of energetic (boosted) DM may also arise from decays/annihilation ~\cite{Agashe:2014yua,Cherry:2015oca, Nagao:2024hit,Bhattacharya:2014yha,Kopp:2015bfa, BetancourtKamenetskaia:2025noa, ICARUS:2024lew, Nagao:2024itk, COSINE-100:2023tcq, Iovine:2023qzn, Aoki:2023tlb, Toma:2021vlw, Giudice:2017zke, Berger:2019ttc, Bhattacharya:2017jaw, Bhattacharya:2016tma}, or through interaction with galactic cosmic rays, extensively studied in Refs. \cite{Bringmann:2018cvk, Ema:2018bih,Cappiello:2019qsw,Dent:2019krz,Guo:2020oum,Dent:2020syp,Bell:2021xff,Xia:2021vbz,Alvey:2019zaa,Wang:2019jtk, Cappiello:2024acu, PandaX-II:2021kai, Maity:2022exk, Reis:2024wfy, Super-Kamiokande:2022ncz, Super-Kamiokande:2017dch,Dutta:2024kuj} and known as cosmic-ray upscattered dark matter (CRDM). Another scenario of interest is supernova-produced DM (SNDM), as studied in \cite{Baracchini:2020owr, DeRocco:2019jti, Das:2024ghw, Xu:2026dae, Jeesun:2026lro,Alonso-Gonzalez:2026syw, Alonso-Gonzalez:2024ems}. SNDM involves a steady-state flux of MeV-scale ``dark matter''\footnote{We use the term ``dark matter'' here in alignment with previous literature on this subject, however, we are not assuming there is necessarily any connection with the particles making up cosmological dark matter halos.} produced in galactic supernovae, which travels semi-relativistically towards Earth.

The Earth's and the solar system's motion through the galactic halo causes DM signals to be strongly anisotropic, resulting in an apparent DM \textit{wind}. Non-relativistic WIMPs from the galactic halo would generate recoil events peaking towards the direction of the star Deneb, in the constellation Cygnus \cite{Spergel:1987kx}. This characteristic signature would enable the unambiguous identification of these signals if the three-dimensional recoil directions could be measured accurately. We will refer to non-relativistic WIMP-like DM as ``halo DM'', to differentiate it from CRDM (halo DM upscattered to relativistic velocities by cosmic rays) and SNDM (semi-relativistic DM produced in supernovae).

CRDM and SNDM also inherit the directionality of their respective astrophysical sources. Assuming cosmic rays are distributed approximately homogeneously through the galactic halo, CRDM will predominantly originate from the direction of the galactic centre (GC), where the DM density is highest. Likewise, the line of sight towards the GC also has the highest column density of past galactic supernovae. So both CRDM and SNDM are expected to have broad angular distributions influenced by the baryonic component of our galaxy, peaking in a direction that is 90 degrees away from the expected directionality of halo DM. Leveraging this directionality to search for types of boosted dark matter has been studied in \cite{Nagao:2022azp, NEWSdm:2023qyb, Baracchini:2020owr, Super-Kamiokande:2022ncz}.

The primary advantage of directionality is that it offers a means to distinguish these models. A halo DM model with a heavy particle mass and an astrophysically-boosted DM model with a light particle can conspire to generate comparable momentum transfers in recoil events. These models would not be distinguishable from one another using nuclear recoil energy signals alone \cite{Baracchini:2020owr}. However, reconstructing their distinct incoming angular distributions would allow the discrimination of these signals in a detector with directional sensitivity.

Conventional DM detectors aim to measure the energy deposited by a scattered nucleus in a target medium. Experiments with directional sensitivity aim to reconstruct the angular distribution of recoil tracks as well. These directional DM experiments are of interest in the field because the angular distribution of events provides one of the only avenues for identifying that a putative DM signal originates from our galaxy~\cite{Hochberg:2021ymx,Nagao:2022azp,NEWSdm:2023qyb,Afek:2021vjy}. The directional signature is important for connecting any putative discovery with the expected astrophysical distribution of DM in our galaxy's halo~\cite{Lee:2012pf,OHare:2014nxd,Kavanagh:2016xfi,OHare:2018trr,OHare:2019qxc}, and, crucially, can be used to distinguish DM events from the otherwise irreducible neutrino background~\cite{Grothaus:2014hja, OHare:2015utx, Mayet:2016zxu, OHare:2017rag, OHare:2020lva}---sometimes referred to as the ``neutrino fog" ~\cite{Billard:2013qya,Ruppin:2014bra,OHare:2016pjy, Dent:2016iht, Dent:2016wor, Gelmini:2018ogy,OHare:2021utq,Akerib:2022ort,Aalbers:2022dzr,Carew:2023qrj,Herrera:2023xun,Maity:2024hzb,XLZD:2024nsu,Dent:2025drd}. This problematic region of parameter space has recently been accessed by large xenon-based dark matter experiments~\cite{LZ:2025igz,XENON:2024hup,XENON:2024ijk,PandaX:2024muv}. However, since these detectors do not measure directionality, it is unclear whether the events reported by these collaborations are truly solar neutrino events or if they are contaminated by true dark matter events. This further motivates the development of directional detectors, a goal pursued by the CYGNUS Consortium \cite{Vahsen:2020pzb,OHare:2022jnx,Vahsen:2021gnb} and the CYGNO collaboration \cite{Baracchini:2020btb,Amaro:2023dxb,Almeida:2023tgn,CYGNO:2023gud,Amaro:2025pms}. 

A promising technology for directional detection is to use gas time-projection chambers (TPCs). In these detectors,  an incoming particle scatters off a target nucleus, producing a recoiling nucleus that leaves an ionisation track. The liberated charge is then transported towards the readout by an electric field. During this drift, the track diffuses to form a three-dimensional ionisation cloud; its projection onto the readout is reconstructed in 2D, while the third dimension is inferred from timing information. Cameras may also be employed in addition to the readout to obtain additional simultaneous recoil information, as is the case for the CYGNO collaboration. The use of a low-density target medium is essential to enable the track to be resolved. However, it has been shown that good tracking for $\gtrsim $~keV scale nuclear recoils is possible even at gas pressures around one atmosphere~\cite{Vahsen:2020pzb,Vahsen:2021gnb}. A wide variety of gas mixtures is possible, providing a high level of flexibility and the ability to tune the target medium to specific DM models and couplings. For concreteness, we focus here on fluorine-based gas mixtures.

In this paper, we examine a scenario in which three dark matter models (halo DM, CRDM and SNDM), which otherwise have indistinguishable rates and energy spectra, could be distinguished by a direction-sensitive detector. We base our envisioned detector on configurations proposed for CYGNUS~\cite{Vahsen:2020pzb}, and study the extent to which different DM scenarios can be distinguished as a function of the detector's performance (angular resolution, energy resolution, and head/tail recognition).

The paper is organised as follows: we describe the directional dark matter signals we will study in Sec.~\ref{sec:dirn_dm}, present our analysis in Sec.~\ref{sec:analysis}, and conclude in Sec.~\ref{sec:conclusions}.

\section{Directional Dark Matter Signals}
\label{sec:dirn_dm}

We start by describing the theoretical calculation of the direction and energy-dependent event rate for each DM model: halo DM, CRDM and SNDM. 

\subsection{Halo Dark Matter}
We assume the commonly used Standard Halo Model (SHM)~\cite{Evans:2018bqy}, where the DM particles are distributed in an isothermal, isotropic sphere with a Gaussian velocity distribution. We consider as the observable signal in a direct detection experiment, the elastic scattering of halo DM with a nucleus in the detector. The motion of the solar system in a circular orbit around the Milky Way introduces a strong anisotropy in the flux of dark matter incident on Earth, which has a peak direction pointing back towards the Cygnus constellation \cite{Spergel:1987kx}. The relative motion of the Earth with respect to the Sun also leads to a temporal modulation in the speed of the lab in the galactic frame, $v_{\textrm{lab}}(t)$. The average of this temporal dependence, $v_{\mathrm{lab}}=232\,\mathrm{km/s}$ is used in the recoil rate below.             

For the doubly differential halo DM recoil spectrum, we use the expression given in Refs.~\cite{Lewin:1995rx,Gondolo:2002np}:
\begin{align}
    \frac{\mathrm{d}^2R}{\mathrm{d}E_{\mathrm{r}}{\mathrm{d}\Omega}}= \frac{\mathrm{d}\sigma_{\mathrm{NR}}}{\mathrm{d}E_r}  \nonumber\\
    \times &\left\{
    \mathrm{exp}\left[-\frac{1}{v_p^2}\left(\frac{\sqrt{2m_T E_r}}{2\mu} - 
    v_{\mathrm{lab}} \,\mathrm{cos}\gamma\right)^2\right] \right. \nonumber\\ 
    &-\left.
    \mathrm{exp}\left[-\frac{v_{\mathrm{esc}}^2}{v_p^2}\right] \right\} \nonumber\\
    \times &\Theta \left(\mathrm{cos}\gamma - \frac{-\left(\frac{\sqrt{2m_T E_r}}{2\mu} - v_{\mathrm{esc}}\right)}{v_{\mathrm{lab}}} \right).   
    \label{eq:NR_rate}
\end{align}
Here, $m_T, E_r, \mu$ are the target mass, recoil energy and DM-target reduced mass respectively, $v_{\mathrm{esc}}= 544~ \mathrm{km/s}$ is the galactic escape velocity, $v_p = \mathrm{220\,\mathrm{km/s}}$ is the value of the circular speed in the SHM's potential, and $\gamma$ is the angle between the direction of nuclear recoil and $-v_{\mathrm{lab}}$, such that 
\begin{align}
    \mathrm{cos}\,\gamma =\mathrm{cos}\,b_r\, \mathrm{cos}
\left(\frac{\pi}{2}-l_r\right) ,
\end{align}
where $(l_r, b_r)$ are galactic longitude and latitude in the laboratory frame. Assuming a spin-independent interaction mediated by a new vector particle with mass $m_A$, the scattering cross section, $\textrm{d}\sigma_{\rm  NR}/\textrm{d}E_R$, is
\begin{align}
    \frac{\mathrm{d}\sigma_{\mathrm{NR}}}{\mathrm{d}E_r} = ~
    &F^2\left(q^2\right) \nonumber\\
    &\times\frac{g_{\chi}^2 g_i^2}{4\pi}\frac{2m_T m_\chi^2-E_r\left((m_\chi+m_T)^2 + m_T E_r^{2}\right)}{(2m_T E_r + m_A^2)^2 \,  m_\chi ^2 v^2}
    \label{eq: nr xsec}.
\end{align}
Here, $m_\chi$ is the DM mass, $v$ its velocity, and we have taken the non-relativistic limit in setting its kinetic energy $T_\chi \rightarrow \frac{1}{2}m_\chi v^2$. The couplings of the vector mediator to the DM and the nucleons are given by $g_\chi$ and $g_i$, respectively. 
For the nuclear form factor $F^2(q^2)$, we adopt the standard Helm ansatz~\cite{Helm:1956zz}.


\begin{figure*}[t]
    \includegraphics[width=1\linewidth]{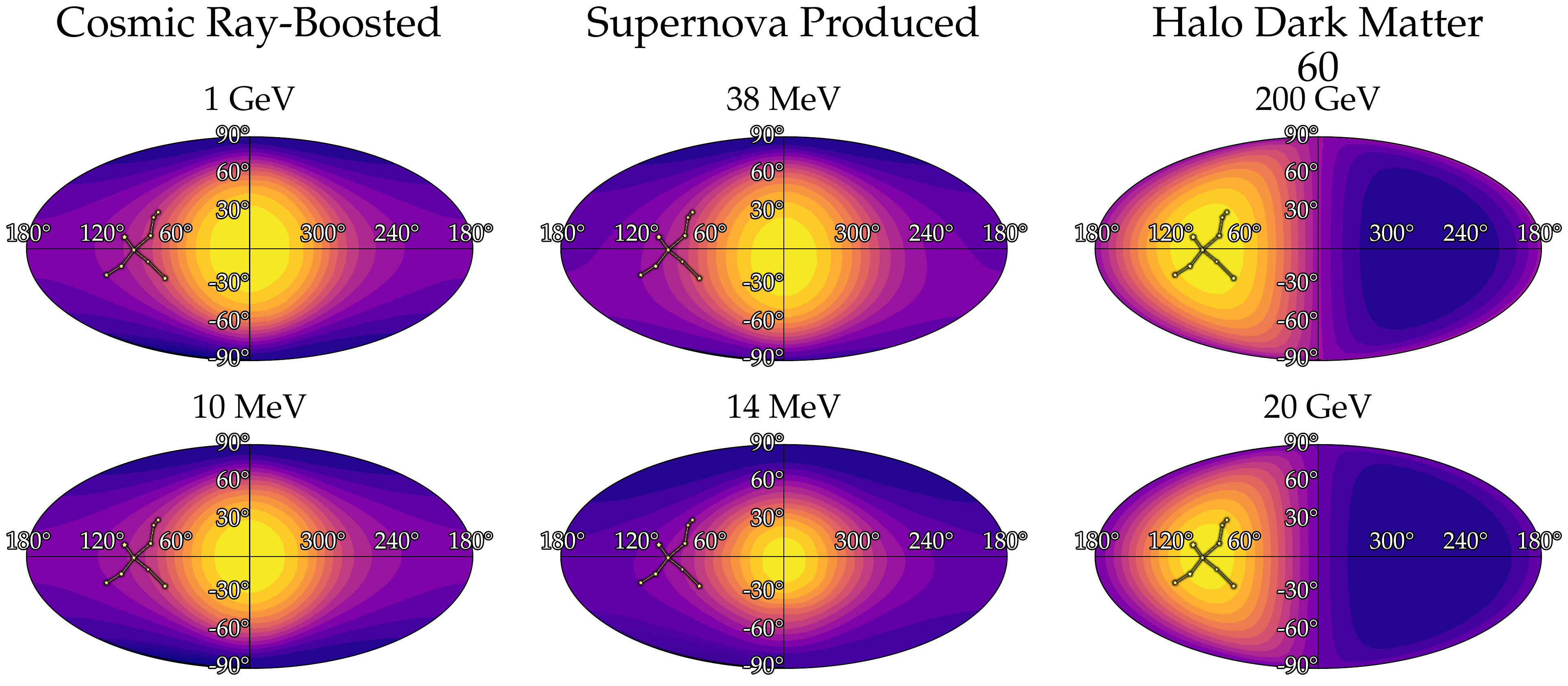}
    \caption{The directional event rates from cosmic ray boosted dark matter (\textit{left}), supernova-produced dark matter (\textit{middle}), and halo dark matter (\textit{right}), each for two different DM mass values. The rates are displayed in galactic coordinates, with the galactic plane running horizontally. }
    \label{fig:mollweide_maps_v2}
\end{figure*}

\subsection{Cosmic-ray upscattered DM}
Experimental sensitivity to CRDM is generally restricted to large interaction strengths because the signal is controlled by two factors of the scattering cross section: the first in the boosting process and the second in the detection process. The relevant cross sections are therefore so large that bounds from perturbativity and unitarity become an important consideration~\cite{Alvey:2022pad, Digman:2019wdm}. Employing light mediators allows for long-range interactions that can exceed the geometric limit imposed by contact interactions~\cite{Digman:2019wdm}, and are sometimes required for probing phenomenologically interesting cross sections. Importantly, light mediators themselves are subject to constraints from astrophysical observations and collider searches, and in many cases the parameter space is already strongly constrained \cite{Bell:2023sdq}. Ref.~\cite{Bell:2023sdq, Dutta:2024kuj} demonstrated that the vector mediator is the least constrained of the light mediator scenarios, and hence it is the model we focus on here. Finally, it is important to note that the interactions between the cosmic rays and halo dark matter involve large momentum transfers ($q^2 \gg m_{\textrm{med}}^2$), hence it is essential to retain the full energy-dependent form of the scattering cross section~\cite{Dent:2019krz}. 

To calculate the CRDM event rates, we describe the DM-nucleon interaction with a vector mediator, which gives rise to spin-independent scattering. The effective Lagrangian that couples the mediator $V_\mu$ to nucleons, $N$, and a fermionic DM candidate, $\chi$, takes the form
\begin{align}\label{eq:Lagrangian}
    \mathcal{L}_{\mathrm{int}}^{\mathrm{vector}} 
    & =  g_{\chi V} V^{\mu} \Bar{\chi}\gamma_{\mu}\chi
    +g_{NV} V^{\mu} \Bar{N}\gamma_{\mu}N\,,
\end{align}
which results in a scattering cross section given by,
\begin{align}\label{eq:cr xsec}
    &\left(\frac{\mathrm{d}\sigma_{it}}{\mathrm{d}T_t}\right)_{\text{vector}}
    =  G^2\left(q^2\right) \times g_{t V}^2 g_{iV}^2A_i^2 \nonumber\\
     &\times\frac{ \left(2m_{t}\left(m_i + T_i\right)^2 - \left(\left(m_i + m_t\right)^2+2m_t T_i\right)T_t + m_t T_t^2\right) }
    {{4\pi\left(T_i^2+2m_{i}T_{i}\right)\left(m_{V}^2+2m_{t}T_{t}\right)^2}}\,.
\end{align}
Here, the $g_{xx}$ are dimensionless coupling constants, $m_i$ and $m_t$ represent the incident and target masses respectively, while $T_i$ and $T_t$ are the incoming and outgoing kinetic energies. For the boosting interaction, the CR is the incident particle and $\chi$ is the target. For the detection interaction, the $\chi$ becomes the incident particle and the target is a nucleus in the detector. For proton and helium scattering, the form factor $G^2 (q^2)$ takes the dipole form with mass $\Lambda_{\text{p}} = 0.770~\text{GeV}\, (\Lambda_{\text{He}} = 0.410~\text{GeV})$~\cite{Perdrisat:2006hj}. For all heavier nuclei, we take the Helm form factor as in Eq.~(\ref{eq: nr xsec}). This is assumed for all the models we consider.

The interaction of Eq.~(\ref{eq:Lagrangian}) enables cosmic rays propagating in the galaxy to scatter off, and boost, non-relativistic dark matter in the galactic halo, producing a flux of relativistic CRDM. The differential flux of the CRDM population at Earth is then,
\begin{equation}
    \frac{\mathrm{d}\Phi_\chi}{\mathrm{d}T_\chi\mathrm{d}\Omega} = 
     \int_{\text{los}} \mathrm{d}s\frac{\mathrm{cos}\,b }{4\pi} \, \frac{\rho_\chi(s)}{m_\chi} \, 
    \int_{T_i^{\textrm{min}}}^{\infty} \mathrm{d}T_i\,\frac{\mathrm{d}\Phi_i}{\mathrm{d}T_i}\frac{\mathrm{d}\sigma_{\chi i}}{\mathrm{d}T_\chi}\,,
\end{equation}
where the integral is taken along the line-of-sight, from the Earth to a distance,  $s$, through the galactic halo.
The integration limit $T_i^{\textrm{min}}$ is the minimum kinetic energy required by a CR particle to upscatter a DM particle to kinetic energy $T_\chi$. In the galactic rest frame, this is
\begin{align}\label{eq:TiMin}
    T_{i}^{\textrm{min}} =
    &\frac{T_\chi-2m_i}{2} \nonumber\\
    +
    &\frac{1}{2}\sqrt{\frac{T_\chi(2m_\chi+T_\chi)(4m_i^2+2m_\chi T_\chi)}{2m_\chi T_\chi}}.
\end{align}

We use an NFW profile for the galactic DM distribution \cite{1996ApJ...462..563N, Ackermann_2012}, in the form 

\begin{align}
\label{eq:NFW}
    \rho_\chi(r) = \rho_{\textrm{NFW}}(r) \equiv \rho_{\textrm{E}}\left(1+ \frac{r_\odot}{R_s}\right)^2 \frac{1}{\frac{r}{r_\odot}\left(1+\frac{r}{R_s}\right)^2}\,,
\end{align}
assuming the Milky Way has a scale radius $R_s = 20$ kpc, and $r_\odot= 8.2~\kpc$. Following Refs \cite{Ema:2018bih, Guo:2020oum, Cappiello:2019qsw} we assume a homogeneous CR density inside a cylinder of radius $R=10$ kpc, centred on the galactic centre, and height $z=2$ kpc, i.e.~ 1~kpc above and below the galactic plane.  We assume the cosmic-ray density is zero outside this region and take the line of sight integral out to a distance of $s = 18.6$ kpc from Earth to cover this cylinder. This ensures a correct computation of the directional signal. In previous work, the integral was truncated at 1 kpc for conservative limit setting \cite{Bringmann:2018cvk}, however that would lead to an artificial directional distribution in this case. 

We use galactic longitude and latitude $(l , \, b)$ such that $\mathrm{d}\Omega = \mathrm{d}l \,\mathrm{d}\cos b$, related to galactocentric cylindrical coordinates by
\begin{equation}
    r(l,b)=\sqrt{r_{\odot}^{2} + s^{2}-2 r_{\odot}(s \cos b \cos l)} \, .
\label{eq:radiusGC}
\end{equation}
The differential recoil rate is then,
\begin{equation}
    \frac{\mathrm{d}^2R}{\mathrm{d}E_r \mathrm{d}\Omega} = 
    \int\mathrm{d}T_\chi \, \mathrm{cos}\,\theta_r \frac{\mathrm{d}\sigma_{\chi i}}{\mathrm{d}E_r} \frac{\mathrm{d}\Phi_\chi}{\mathrm{d}T_\chi\mathrm{d}\Omega}\,,
\end{equation}
where
\begin{equation}
    \mathrm{cos}\,\theta_r = \left(\frac{E_r(m_\chi+m_T+T_\chi)}{T_\chi\, (2m_\chi+T_\chi) (2 m_T +E_r)}\right)^{1/2},
\end{equation}
and $\theta_r$ is the angle between the direction of the incoming dark matter particle and the recoiling nucleus. The event rate for a given scattering angle is assumed to be azimuthally symmetric around the incoming dark matter direction.


\subsection{Supernova DM}
\label{sec:iiic}
Supernova-produced DM (SNDM), discussed in Refs.~\cite{DeRocco:2019jti, Baracchini:2020owr}, refers to a flux of MeV-scale particles generated during galactic supernovae (SN), which then travel semi-relativistically throughout the Galaxy. Due to the semi-relativistic speeds, the burst from a given SN arrives at Earth with a spread of arrival times of order 100-1000 years.  As a result, the bursts from many past SN overlap, producing a 
flux at Earth that is approximately---although not precisely~\cite{Alonso-Gonzalez:2026syw}---constant in time. Core collapse supernovae can efficiently produce dark-sector particles with masses up to their temperature, $T\sim 30\,\textrm{MeV}$.
The directionality of the flux will peak towards the galactic centre, which is the line-of-sight where the column density of stars, and hence past supernovae, is largest. 

Following Ref.~\cite{Baracchini:2020owr}, we take the SN rate density to be
\begin{equation}
    \frac{\mathrm{d} n_{\rm S N}}{\mathrm{d} t}  = A e^{-R /R_0} e^{-|z| /z_0} \, ,
\end{equation}
where $z$ is the height above the galactic plane and $R$ is the galactocentric radius. We take the scale radius and scale height for Type II SN from Ref.~\cite{Adams:2013ana}: $R_0 = 2.9$~kpc and $z_0 = 95$~pc. We normalise the rate density such that the integral of $\mathrm{d}n_{\rm SN}/\mathrm{d}t$ over the galaxy gives two supernovae per century, i.e.~$A = 0.00208$ kpc$^{-3}$ yr$^{-1}$. If each SN generates $N_\chi$ particles, then the flux is simply given by the line-of-sight integral over the rate density,
\begin{equation}
    \frac{\mathrm{d}\Phi}{\mathrm{d}\Omega} = \frac{N_\chi}{4\pi} \int_0^\infty \frac{\mathrm{d} n_{\rm S N}}{\mathrm{d} t} \mathrm{d}s \,.
\end{equation} The distance along the line of sight $s$ is related to the radius and height of the galactocentric cylindrical coordinates by Eq.~(\ref{eq:radiusGC}), 
and
\begin{equation}
\label{eq:galactocentric height}
    z = z_\odot + s \sin b ,
\end{equation}
where the Earth position is $z_\odot= 24~{\rm pc}, r_\odot= 8.2~{\rm kpc}$ as before.

The SNDM momenta, $p$, as measured at Earth, follow a Fermi-Dirac distribution,
\begin{align}
    f_\ast(p) = \left[\mathrm{exp}\left(
    \frac{1}{T}
    \frac{\sqrt{p^2 + m_\chi^2}}{1-2\Phi}\right)+1\right]^{-\frac{1}{2}}\, ,
\end{align}
where the temperature $T$ is set by the last radius at which the SNDM is thermally coupled to the standard model (SM) (the energy sphere, $r_E$), and $\Phi \equiv \int_{r_E}^\infty \textrm{d}r \,m_{\mathrm{enc}}(r)/r$ \cite{Baracchini:2020owr} is the associated redshift factor.

We only consider a heavy vector mediator with mass $m_V = 1~\mathrm{GeV}$, and hence take the cross section to be constant~\cite{Baracchini:2020owr},
\begin{equation}
\label{eq:SNDM xsec xn}
    \sigma_{\mathrm{SNDM}, \chi n} =
    \left(\frac{4\mu^2y}{m_\chi^4}\right)
    \left(\frac{\Lambda}{m_V}\right)^4\, ,
\end{equation}
where,
\begin{align}
    y = \frac{\epsilon^2 g_d^2}{4\pi} \left(\frac{m_\chi}{\Lambda}\right)^4\, .
\end{align}
Here, $\epsilon$ and $g_d$ are the nucleon coupling to the mediator and dark charge respectively, and we take $\Lambda = 1\,\textrm{GeV}$ such that the mediator is heavy with respect to momentum transfer, and hence it is valid to express the strength of this interaction in terms of the effective parameters $y$ and  $\Lambda$. The nuclear scattering cross section is then related to Eq.~(\ref{eq:SNDM xsec xn}) by,
\begin{align}
    \sigma_{\mathrm{SNDM},\chi N} = A^2 \frac{\mu_{\chi N}^2}{\mu_{\chi n}^2}\times\sigma_{\mathrm{SNDM},\chi n}.
\end{align}

From \cite{Dho:2023ibh}, the double differential spectrum of the nuclear recoils induced by SNDM follows:
\begin{align}
    \frac{\mathrm{d}R}{\mathrm{d}E_r\,{\mathrm{d\,cos}\,\theta}} 
    \propto& 
    \frac{1}{(1-2\Phi)^{3/2}}  \sigma_{\mathrm{SNDM},\chi N}
     \nonumber\\ 
    \times&\int \mathrm{d}p\, F^2(q^2)\,\delta\bigg(\mathrm{cos}\,\theta - \frac{\sqrt{2m_T E_r}}{2p}\bigg)   \nonumber\\ 
    \times& \sqrt{\frac{p^2+2\Phi m_\chi^2}{p^2 + m_\chi^2} }f_\ast(p)\, .
\end{align}
We will choose the benchmark values for $T$, $\Phi$ for the SNDM cases according to scenarios 5 and 6 in Ref.~\cite{Baracchini:2020owr} when we compare these rates with the CRDM and halo DM cases.

\begin{figure}[t]
    \centering
    \includegraphics[width=1\linewidth]{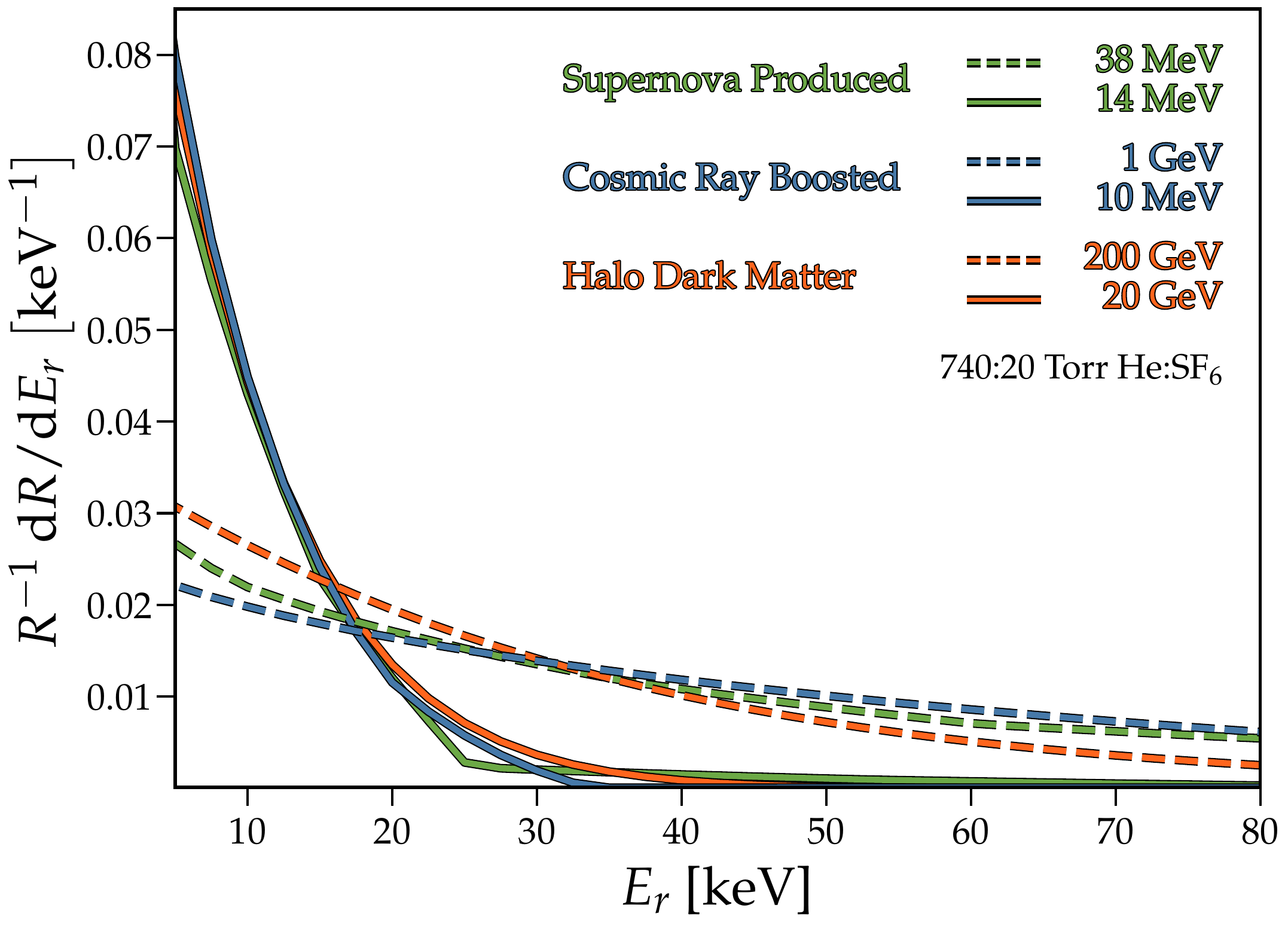}
    \caption{Recoil spectra for all DM models. The spectra extend to much higher energies for the high-mass cases than for the low-mass cases, as expected kinematically. The three rates for each set (solid and dashed lines) are normalised such that they generate the same number of events above a threshold of 5 keV.}
    \label{fig:drde_allmodels}
\end{figure}

\begin{table}[t]
    \centering
    \caption{Benchmark models and parameters
    }
    \begin{tabular}{c|c c c c c}
    \hline \hline
    Model & DM mass & T(MeV)\footnotemark[1] & $\Phi$\footnotemark[1] &  $y^{\textrm{ref}}$~\footnotemark[1] & $\sigma_{\chi p}^{\textrm{ref}}~(\textrm{cm}^2)$ \footnotemark[2]\\
    \hline
    CRDM & 10 MeV & - & - & - & $7 \times 10^{-29}$ \\
     & 1 GeV & - & - & - & $10^{-32}$ \\
    \hline
    SNDM & 14 MeV & 3 & 0.07 & $10^{-15}$ & $10^{-37}$\\
    & 38 MeV & 13.4 & 0.1 & $10^{-16}$ & $10^{-39}$ \\
    \hline
    Halo DM & 20 GeV & - & - & - & $2 \times 10^{-48}$ \\
    & 200 GeV & - & - & - & $3 \times 10^{-48}$ \\
    \hline \hline
    \end{tabular}

    \footnotetext[1]{Only relevant for SNDM.}
    \footnotetext[2]{Cross sections in regions of parameter space not yet excluded by LZ for the DM masses considered \cite{LZ:2024zvo,Bell:2023sdq, DeRocco:2019jti}.}
    \label{tab:cases}
\end{table}

\begin{figure*}[!t]
    \centering
    \includegraphics[width=1\linewidth]{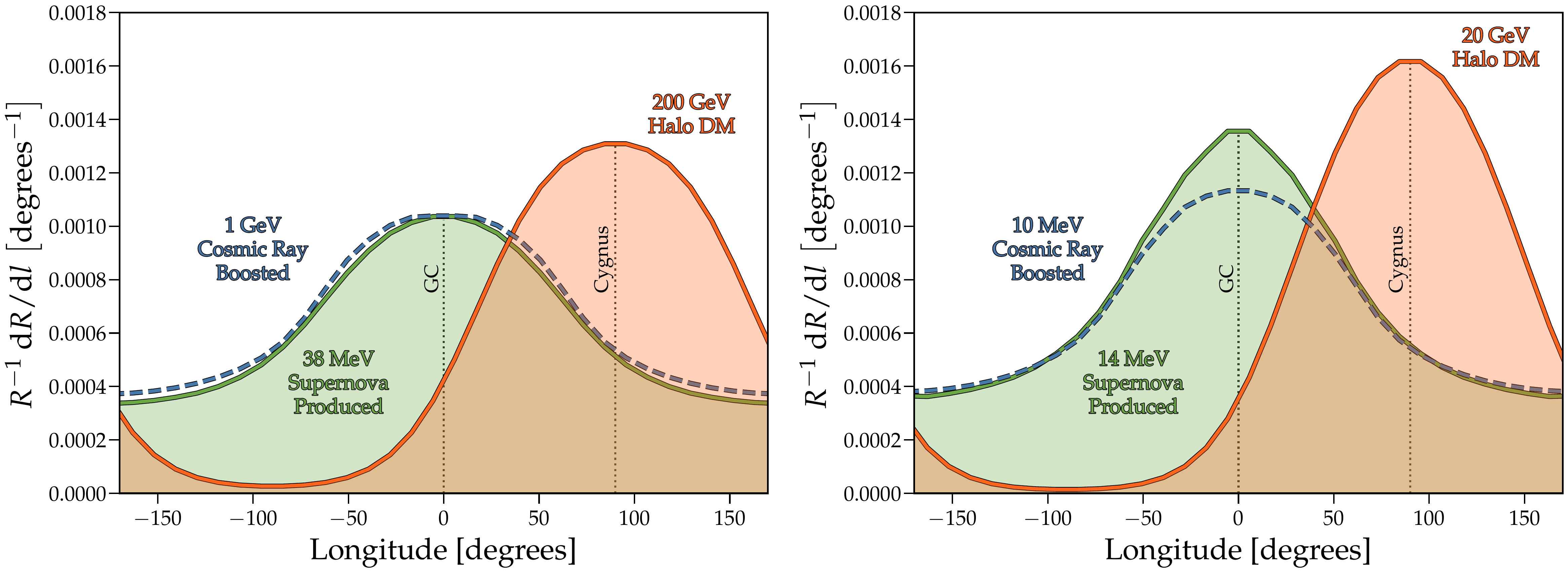}
    \caption{The directionality of normalised scattering rate with galactic longitude, for high mass (\textit{left}) and low mass (\textit{right}) CRDM, SNDM and halo DM at fixed galactic latitude.  The angular spectra for the choice of SNDM and CRDM parameters demonstrate very similar shapes, as expected, with peaks lining up at the galactic centre $l_{\textrm{GC}} = 0\degree$. The halo DM peak exhibits a perpendicular arrival direction, coming from $l_{\textrm{Cygnus}} \sim 90\degree$.}
    \label{fig:combined_dirn_rate}
\end{figure*}


\subsection{Comparison}
\label{subsec:Comparison}

We now demonstrate that the halo DM, SNDM and CRDM models can be challenging to distinguish using the energy spectra of their nuclear recoils alone.  This occurs because there are  parameter choices for which the spectra become highly degenerate. We choose two scenarios to illustrate this idea, consisting of two sets of models generating low-energy spectra and high-energy spectra. The parameters are listed in Table~\ref{tab:cases} and the differential recoil energy spectra are shown in Fig.~\ref{fig:drde_allmodels} for a 740:20 Torr ratio He:$\textrm{SF}_{6}$ gas mixture. The two sets of recoil spectra are highly similar to one another. In a real experiment with limited statistics and subject to limitations in the precision with which $E_r$ could be reconstructed, these models would be indistinguishable.

In contrast to this, the three classes of models generate very different angular distributions. The production of CRDM occurs primarily in regions of high dark matter density, so its angular recoil spectrum peaks towards the direction of the galactic centre. Similarly, we expect the SNDM case to also peak towards the GC because the rate density of supernovae correlates with the stellar distribution in the galaxy. On the other hand, the halo DM signal is unrelated to the distribution of DM outside of the solar system and is uncorrelated with the baryons. The directionality of the incident flux of DM in this case originates solely from our motion through the halo, which is perpendicular to the galactic centre. The three angular distributions integrated over recoil energy are shown side by side in Fig.~\ref{fig:mollweide_maps_v2} for both the low-mass and high-mass sets of models.

The halo DM angular distribution spans effectively a single hemisphere of the sky centred on the Cygnus constellation. The SNDM and CRDM distributions in contrast are peaked towards the galactic centre---where the profile of both the supernova rate and DM density profile are sharply rising\footnote{We note that the SNDM distribution peaks at a latitude slightly below the origin in Fig. \ref{fig:mollweide_maps_v2}, which is due to the fact the  Earth sits slightly above the galactic plane by a height given in Eq.~(\ref{eq:galactocentric height}).}---and they have wide tails that wrap around the full range of longitudes due to particles originating from the galactic disk. We see this more clearly in Fig.~\ref{fig:combined_dirn_rate}, where we show a slice through the angular distribution for $b = 0 \degree$ as a function of galactic longitude $l$. The high-mass set of models (left panel of Fig.~\ref{fig:combined_dirn_rate}) exhibits a broader angular distribution than the low-mass set of models (right panel). This is a result of the fixed lower-energy threshold above which we are integrating the rate, i.e.~we are cutting out more of the low-energy/large-scattering-angle tail of the distribution in the right-hand panel than in the left-hand panel due to their differing $\textrm{d}R/\textrm{d}E_r$ spectra.

We also point out that, although the SNDM case is slightly more peaked than the CRDM case, the differences between the two scenarios are very small. This would likely mean that these classes of models would remain indistinguishable unless a large number of events were collected, and these subtle differences in the angular distributions could be measured. The primary goal here then is to illustrate how these boosted DM populations can be distinguished from a halo population, as opposed to how these boosted populations can be distinguished from each other, which would be much more challenging.

In what follows, we will put each model on an even footing by normalising the rates to generate the same total number of events inside the detector. One can then map these numbers to a corresponding cross section and detector exposure using the reference values of the model parameters listed in Table~\ref{tab:cases},
\begin{align}
   N_{\textrm{CRDM}} = (g_\chi g_p)^4\times\frac{\textrm{exposure}}{\textrm{m}^3\,\textrm{yr}} \, ,
   \label{eq:crdm scale}
\end{align}
\begin{align}
    N_{\textrm{SNDM}} = \frac{y}{y^{\textrm{ref}}} \times \frac{\textrm{exposure}}{\textrm{m}^3\,\textrm{yr}} \, ,
    \label{eq:sndm scale}
\end{align}
\begin{align}
    N_{\textrm{halo-DM}} = \frac{\sigma_{\chi p}}{\sigma_{\chi p}^{\textrm{ref}}} \times \frac{\textrm{exposure}}{\textrm{m}^3\,\textrm{yr}} \, .
    \label{eq:halo DM scale}
\end{align}
Note the CRDM scattering rate scales with $(g_\chi g_p)^4$ due to the \textit{two} factors of cross section required to produce one event. We use units of m$^3$-year for the exposure, assuming a fixed gas mixture and pressure: 740:20 Torr He:SF$_6$, which we motivate further in the following section. We emphasise that since this study focuses on the \textit{distributions} of recoil events---which are dominated heavily by fluorine recoils---our results are insensitive to the exact gas mixture. We would obtain almost identical results for a He:CF$_4$ gas mixture, for instance. 


\section{Analysis}
\label{sec:analysis}

Taking inspiration from the CYGNUS and CYGNO experiments, we consider an experiment consisting of a directional gas time-projection chamber (TPC) which is able to achieve event-by-event reconstruction of nuclear recoil track directions and their energies. To perform an analysis of the sensitivity to these models in this class of directional recoil detector, we first generate true recoil events drawn from the halo DM, SNDM, and CRDM double differential rates, then account for detector effects such as angular resolution and head-tail reconstruction, which limit the detector's ability to fully reconstruct the underlying angular distribution. We will then use two statistical methods, detailed in Refs.~\cite{Green:2010zm, OHare:2017rag}, to determine the number of events required to (i) reject isotropy at a given significance level, (ii) measure the median galactic longitude of the angular distribution so as to confirm the DM signal originates from a particular direction.

\subsection{Detector performance}
\label{subsec:perf}
When a DM particle recoils on a nucleus in the gas volume, it first creates an ionisation cloud of mm--cm in length. In a TPC, this ionisation track is then drifted via an applied electric field to a highly-segmented gas amplification and readout place. Modern ``recoil-imaging" TPCs employ highly-segmented micro-pattern gas detector (MPGD)-based readouts which enable sub-mm spatial resolution of this two-dimensional projection of the cloud (see discussions of ``recoil imaging'' in Refs.~\cite{Vahsen:2021gnb, OHare:2022jnx,Surrow:2022ptn}). Assuming knowledge of the drift velocity, sampling the charge arrival times allows the final projection of the track in 3D to be reconstructed, while the total ionisation is correlated with the initial recoil energy, subject to nuclear quenching.

The track axis and vector sense inferred on each recoil will be correlated with the true underlying direction of the recoiling nucleus, however this reconstruction will be limited by the resolution of the detector and the size of the initial track compared to the diffusion scale and the readout segmentation. Reducing the limitations due to diffusion is of critical importance for DM searches which benefit from higher gas densities (to enhance the event rate), but which reduce the directional performance by increasing the diffusion. DM signals are also exponentially falling with energy, and low-energy tracks are shorter than high-energy tracks, meaning directionality can be lost altogether as a signal at energies well above the energy threshold for detecting events. The choice of gas mixture and its pressure must be carefully optimised due to these considerations.

The CYGNO collaboration at Laboratori Nazionali del Gran Sasso (LNGS) is currently employing a 60\% Helium and 40\% CF$_4$ mixture, favouring the choice of a scintillating gas such as CF$_4$ which allows the use of an optical readout (cameras)~\cite{Torelli:2024mof}. Another possibility is to take advantage of negative ion drift (NID), where an electronegative gas such as SF$_6$ is added to the mixture; this captures the primary ionisation electrons, resulting in the formation of negative ions, which are transported to the readout instead of the electrons. The heavier charge carriers used in NID mean that diffusion can be greatly reduced~\cite{Martoff:2000wi, Marques:2025ahu}, which is an essential advantage for the short tracks foreseen to be generated by DM. A helium and SF$_6$ mixture is well-motivated, because SF$_6$ is a NID gas and the addition of helium provides additional higher energy and longer-track recoil events, while being low enough in density that its addition does not reduce the directionality of the fluorine recoils~\cite{Vahsen:2020pzb}. Helium also provides access to lower DM masses than fluorine due to the fact its recoil spectrum for a fixed DM mass extends to higher energies.

We choose a gas mixture consisting of 740 Torr of helium and 20 Torr of SF$_6$ (bringing the total mixture to atmospheric pressure). This mixture has been shown in previous gas simulation studies to maintain good directionality over the nuclear recoil energies of the signals we consider in this work~\cite{Vahsen:2020pzb}. Other configurations, such as hybrids between the two mixtures, are also being considered in the community~\cite{Marques:2025ahu}, so we choose this mixture simply for concreteness. We emphasise that the details of the gas mixture do not directly influence our results in a significant way because we have chosen models which generate similar recoil spectra and will normalise each event rate to give the same number of signal events. Instead, it is the parameterisation of the detector's performance that influences our results, which is indirectly related to the gas mixture (i.e., high gas densities would have poor performance and low gas densities better performance). This means our results can be readily interpreted for a wide range of potential mixtures, assuming the performance benchmarks (discussed below) can be achieved. Our performance benchmarks are inspired by the different charge readout choices studied in Ref.~\cite{Vahsen:2020pzb}.

The performance limitations we consider are: angular resolution, energy resolution, and head/tail recognition (i.e.~reconstructing the vector sense of $\mathbf{q}$). These are implemented using a Monte Carlo procedure detailed in Refs.~\cite{Lisotti:2024fco, Shekar:2025xhx,OHare:2026pxf}, which we briefly summarise here. We start by generating events with recoil direction unit vectors $\hat{\textbf{q}}$ and energies $E_r$ from each of the double differential rates. We specify $\hat{\textbf{q}}$ in Galactic coordinates, which means we are implicitly assuming a detector which can reconstruct the track in three dimensions (as opposed to some classes of experiment which can only reconstruct a one or two-dimensional projection~\cite{Vahsen:2021gnb,OHare:2017rag}). We assume a simplistic event-level detector efficiency by discarding all events below the threshold, which we set optimistically to E$_{\rm th} = 5$ keV. Since the performance parameterisations are ultimately inspired by detector simulations, they are quoted in terms of true recoil energy and so have information about the nuclear quenching factor folded in. We note here again that our results are not heavily dependent on the choice of recoil energy threshold for the same reasons that they are not very sensitive to the gas mixture, as discussed above.

\subsubsection{Angular Resolution}

We model the finite angular resolution by taking each simulated direction $\hat{\textbf{q}}$ and then sampling a new mis-measured direction $\hat{\textbf{q}}'$ from a von Mises-Fisher distribution centred on $\hat{\textbf{q}}$ which has the following form,
\begin{equation}
    K(\hat{\textbf{q}},\hat{\textbf{q}}',E_r) =  \frac{\kappa(E_r)}{4 \pi \sinh \kappa (E_r)} \exp(\kappa(E_r) \textbf{q}\cdot \textbf{q}'),
    \label{eq: vmf}
\end{equation}
where the parameter $\kappa$ controls the concentration of the distribution around the true direction. Here we will not consider $\kappa$, but rather the more familiar root-mean-squared angular resolution $\sigma_\theta$, which is nevertheless a monotonic function of $\kappa$ that can be evaluated numerically. We parameterise the recoil-energy dependence of the angular resolution as follows,
\begin{equation}
    \sigma_\theta = p_\theta \sqrt\frac{E_{\rm ref}}
    {E_r}.
\end{equation}
Here, $E_{\rm{ref}}$ is an arbitrary reference energy which we take to be 42.5 keV; this corresponds to the 80th and 99th percentile of the recoil energy spectra for the higher-mass and lower-mass halo DM models respectively. At this energy, we take $p_\theta$ to be equal to 20$^\circ$, meaning 80\% and 99\% of the corresponding spectra will be observed at least at this resolution. The choice of this value is informed by the `postdrift' angular resolution obtained in simulations for CYGNUS, with our model closely approximating the experimental performance at energies above the threshold \cite{Vahsen:2020pzb}. The postdrift scenario is what we define as our ``ideal" benchmark, corresponding to a best-case scenario reconstruction of the ionisation cloud that is only limited by the charge diffusion over the drift distance of 25--50~cm in this gas mixture. This could be achieved by any low-noise readout whose segmentation was much smaller than the diffusion scale, e.g.~a pixel readout. The other two performances we consider, namely ``realistic" and ``poor", are taken as the performances achievable by a strip readout in the latter case ($p_\theta = 30^\circ$), and an intermediate performance between the two in the former ($p_\theta = 25^\circ$). The strip readout would have a coarser-grained resolution but would be much cheaper than a pixel readout, so our performance benchmarks are all `realistic' in the sense that they could be achieved in a real experiment, but the cost of doing so would be very different. We refer to Ref.~\cite{Vahsen:2020pzb} for further discussion.

\subsubsection{Head-Tail Effect}

\begin{figure*}[!t]
    \centering
    \includegraphics[width=1\linewidth]{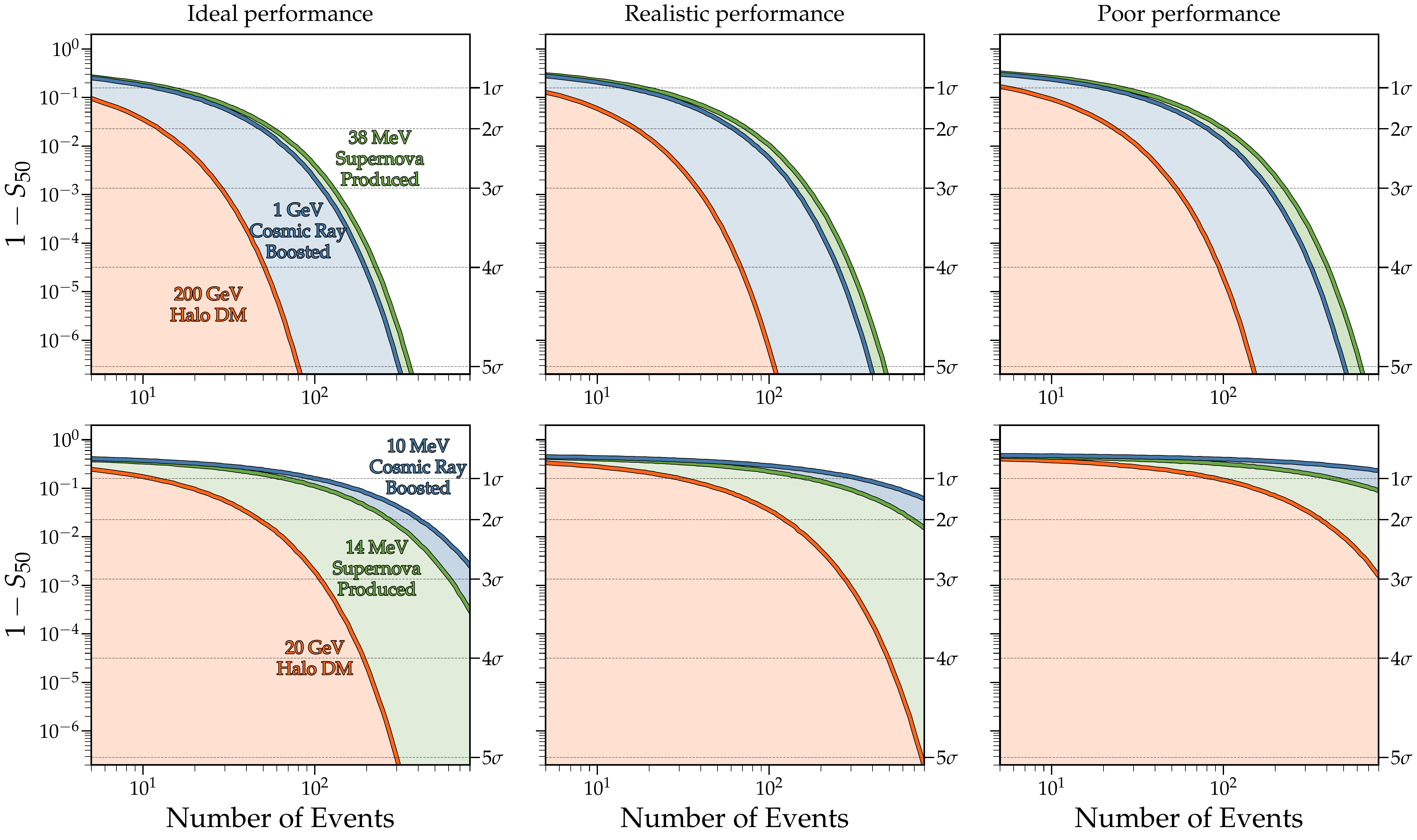}
    \caption{The median significance, $S_{50}$, at which an isotropic background could be rejected as a function of the number of detected signal events. The top three panels show the higher-mass halo DM, SNDM and CRDM models, while the bottom panels show the lower-mass cases. Detector performance degrades from left to right.}
    \label{fig:1-50}
\end{figure*}

We next consider how well the detector can identify the sense of the recoil track, which is distinct from inferring the track's \textit{axis} (which is handled by our angular resolution smearing in the previous section). This is measurable in practice by measuring the `head' and `tail' of the recoil's ionisation distribution. For nuclear recoils of $\gtrsim$keV energies, the $\textrm{d}E/\textrm{d}x$ profile declines along the track as the recoil slows down. We model the ability to correctly assign the track's vector sense by either maintaining or flipping the sign of each generated vector with a probability sampled from the following energy-dependent function,
\begin{equation}
    \varepsilon_{\rm HT}(E_r) = \frac{1}{2}\left(\frac{1}{1+e^{-a_{\rm HT}(E_r-E_{\rm HT}))}}+1 \right) \, .
\end{equation}
This is a sigmoid function that begins at 0.5 for lower recoil energies near threshold (where the tracks are so short that the head/tail assignment is effectively random), and then rapidly approaches 1 as the energy increases. High-energy recoil tracks are longer and produce more ionisation, meaning the characteristic $\textrm{d}E/\textrm{d}E_r$ profile is more easily measured. We adopt three performances following the same naming scheme introduced above, where $a_{\rm HT}= 0.2$ and $E_{\rm HT}$ is equal to $16$~keV, $20$~keV and $24$~keV  for the ``ideal", ``realistic" and ``poor" cases respectively. 

\subsection{Rejecting Isotropy}\
\label{subsec:rejecting}

To quantify the extent to which the directionality of each DM signal can be detected, we first determine the number of events needed to discriminate our signals from an isotropic background at some significance level. The result of this test depends on how anisotropic the distributions are and, importantly, can be performed using only a collected sample of events, and does not require modelling their expected angular distributions. Following Refs.~\cite{PhysRevD.72.123501, OHare:2017rag}, we consider the quantity,
\begin{equation}
    \langle \cos \theta_{\rm rec} \rangle = \frac{\sum_{i=1}^N \cos\theta^i_{\rm rec}}{N}
\end{equation}
which is the average of the cosine of the recoil angles between a vector $\hat{\textbf{q}}_i$ and the expected recoil direction $\hat{\textbf{q}}_{\rm rec}$. In galactic coordinates, the latter direction is taken to be equal to $\hat{\textbf{q}}_{\rm Cyg}=[0,1,0]$ in the case of non-relativistic halo dark matter signals and $\hat{\textbf{q}}_{\rm GC}=[1,0,0]$ for signals originating from the Galactic Centre, namely SNDM and CRDM. While the test requires an a priori choice of coordinate system, there is no requirement that this chosen direction needs to align with the correct mean or median direction. The test performs the best when the chosen direction aligns with the directionality of the signal. That said, these two directions would be the only natural ones to choose in the context of DM search.

We generate $\langle \cos \theta_{\rm rec} \rangle$ distributions from $10^5$ Monte Carlo experiments for each model. We test these distributions against an isotropic null hypothesis. Under this hypothesis, the distribution of the test statistic follows a Gaussian with standard deviation $\sigma = \sqrt{\frac{1}{3N}}$ and zero mean. From our Monte Carlo simulated distribution of test statistics we compute $S_{50}$, which we define to be the median significance at which the null hypothesis can be rejected. The value of $S_{50}$ for each model as a function of the number of detected events is shown in Fig.~\ref{fig:1-50}. To aid in visual clarity, what we actually plot is the $p$-value corresponding to the significance, i.e.~lower values of $1-S_{50}$ indicate a stronger rejection of isotropy. Gaussian significance values in units of $\sigma$ are also highlighted (recall that our null hypothesis test statistic distribution is Gaussian by construction).

In the higher-mass recoil scenario, isotropy can be rejected at a given $\sigma$ with fewer recoil events across all three detector performances, despite these presenting a broader angular distribution (Fig.~\ref{fig:combined_dirn_rate}). This is expected from our discussion in Section~\ref{sec:iiic}: the higher-mass recoils are more energetic, and detector performances increase proportionally with energy (Fig.~\ref{fig:drde_allmodels}).   
Furthermore, in this scenario, rejecting an isotropic background to a given significance level under halo DM requires roughly one quarter of the event count needed for the other models, which holds across all three performance benchmarks. This is consistent with the sharper halo DM median longitude distribution shown in Fig.~\ref{fig:combined_dirn_rate} compared to SNDM and CRDM. The latter two signals largely overlap in this figure, consistent with their very similar angular distributions. Under our best-case scenario for the performance of the detector, a $>3\sigma$ detection would require as few as 30 events for halo DM, but many hundreds for other models. We emphasise, however, that this test is fully non-parametric and so does not require anything about the underlying signal to be modelled. So the detection of a non-isotropic signal with only a few events would immediately suggest not only the presence of a DM signal, but of a \textit{halo DM} signal specifically.

\subsection{Resolving the true direction}
\label{subsec:resolving}

\begin{figure*}[!t]
    \centering
    \includegraphics[width=1\linewidth]{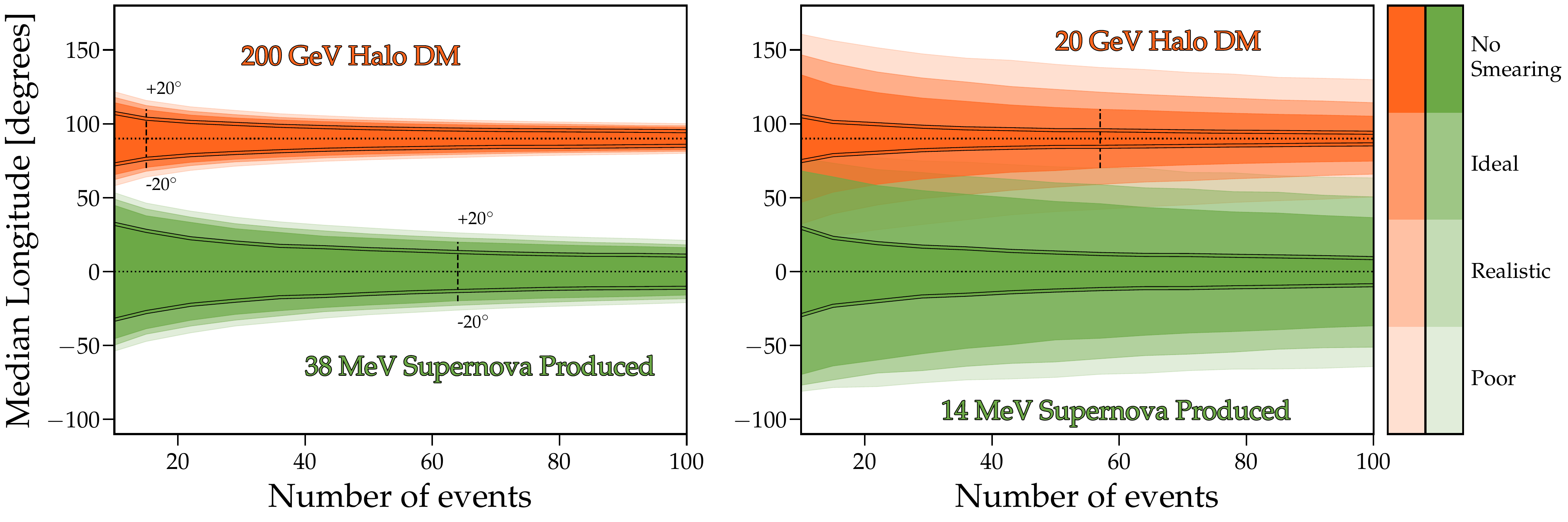}
    \caption{The interquartile ranges for the median reconstructed longitude from $10^5$ Monte Carlo experiments as a function of the number of recoil events. The left panel shows the higher-mass models, while the right shows the lower-mass models. Degrading detector performance is indicated by increasing transparency, with halo DM in green and SNDM in orange. The darkest colour shows the case when we do not include any of the performance limitations discussed in the text, which we use for comparison only. The median direction approaches the true source direction with an increasing number of events, as expected, while the width of the bands quantifies how accurate the reconstructed median direction would be in at least 50\% of hypothetical experiments. We consider the two models to be distinguishable as long as the two regions do not overlap. We omit the CRDM case in this figure as it overlaps with the SNDM case.
    \label{fig:delta_fig}}
\end{figure*}

Our next goal is to establish the extent to which a collection of recoil directions could be used to not only reject the hypothesis of an isotropic background, but provide a confirmation of the underlying directionality, i.e.~DM \textit{discovery}. To this end, we calculate the median direction of recoil events generated from the three distributions and test this median direction against some hypothesised direction: either the direction of Cygnus or the galactic centre. 

The median direction $\hat{\textbf{q}}_{\rm med}$ is the one that minimises the sum
\begin{equation}
    \sum_{i=1}^N \cos^{-1} (\hat{\textbf{q}}_{\rm med} \cdot \hat{\textbf{q}}_{\rm i}),
\end{equation}
where $\hat{\textbf{q}}_{\rm i}$ are the individual recoil vectors and $N$ is the number of events. We compute the direction $\hat{\textbf{q}}_{\rm med}$ for each pseudo-experiment and then extract its Galactic longitude, which is to be compared with the Galactic longitude of either Cygnus ($l = 90^\circ)$ or the Galactic centre ($l = 0^\circ$). The distributions of these median longitudes are shown in Fig.~\ref{fig:delta_fig} as a function of the number of detected events. For each model, and for each performance benchmark, we collapse the distribution down to an interquartile range, i.e.~an interval enclosing 50\% of the distribution centred on the median.

We again examine the two mass scales separately, with the higher-mass recoils on the left and the lower-mass on the right. The \textit{underlying} longitude distribution is wider for the higher mass models as can be seen in the distributions visualised in Figs.~\ref{fig:drde_allmodels},~\ref{fig:combined_dirn_rate}. However, the energy dependence of the detector performance leads to less smearing for these models compared to the low-mass models---the higher mass models generate higher-energy recoils, which have their directions better measured. In this case, the median longitude would be measured to within 20 degrees of the DM's true underlying directionality with as few as $\sim$20 events for the halo dark matter, and $\sim$65 for the signals coming from the galactic centre, assuming an ``ideal" performance. The required event numbers for the lower-mass models make this much more challenging, which means that a true discovery of a boosted DM signal may not be feasible in all cases, unless a large number of events were detected or the signal was modelled in a more detailed way, e.g.~folding in the information about the angular distribution and the recoil energy spectrum into a parameterised model. We emphasise again that (1) we have cherry-picked models which are the \textit{hardest} to distinguish so as to showcase what could be done using directionality; and (2) we are using fully non-parametric hypothesis tests that would not require any modelling of the signal. So, regardless of detector performance, discerning these signals would be completely impossible without access to directional information.

 \begin{figure*}[!t]
    \centering
    \includegraphics[width=1\linewidth]{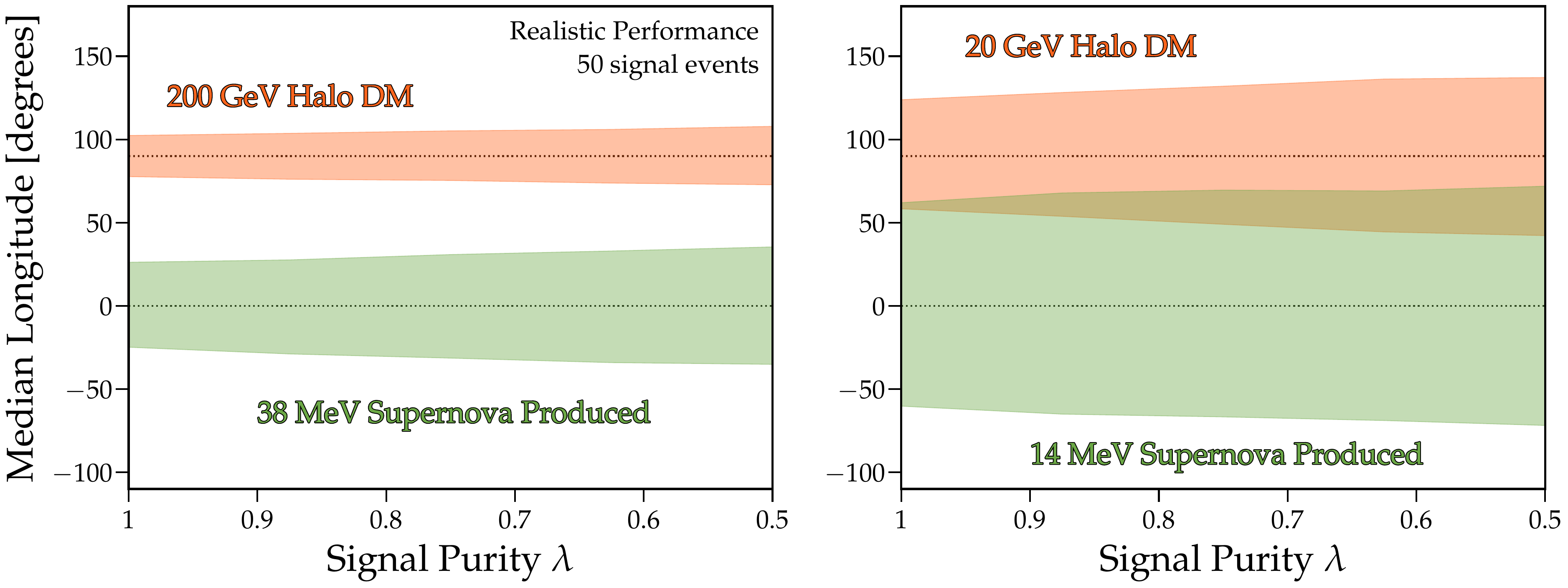}
    \caption{Median longitude interquartile range for the higher-mass (right panel) and lower-mass models (left panel) assuming 50 signal events as a function of decreasing signal purity, $\lambda$. Here, a background-free operation corresponds to $\lambda=1$, while $\lambda=0.5$ means that the detected events are composed of 50\% signal and 50\% background events.}
    \label{fig:sig_pur}
\end{figure*}

\subsection{Signal Purity}
\label{subsec:bg}

Our analysis up to now has been performed assuming close to background-free conditions of the detector could be achieved---as was the case for the DRIFT-II directional DM search~\cite{DRIFT:2014bny}. Although it is true that backgrounds can be severely minimised in a detector like this, it is likely that some background, particularly from electron recoils generated by internal and environmental radioactivity will be present and contaminate the signal. So to conclude our analysis we will repeat our final result but accounting for some non-negligible background contamination. Continuing in the spirit of what we have presented so far, we will not assume that our statistical test has any knowledge of the background's properties (which is a pessimistic assumption) other than it being roughly isotropic. We then only need to quantify the background in terms of a signal purity parameter $\lambda = R_{\rm sig}/(R_{\rm bg}+R_{\rm sig})$.

For brevity, we consider just the ``Realistic" detector performance from Fig.~\ref{fig:delta_fig}. We fix the signal event count to be 50 and vary the number of background events $50 \times (1-\lambda)$. In Fig.~\ref{fig:sig_pur} we show the result for the interquartile range of the median longitude distribution, using the same format as the previous figure, only now as a function of $\lambda$. The background-free case ($\lambda = 1$) corresponds to a slice of Fig.~\ref{fig:delta_fig} at 50 events

As the background level increases ($\lambda$ decreases), the band widens as expected because the median direction becomes less accurate due to the contamination by a sample of isotropically distributed events. The higher-mass case is less affected by the increasing background level: even considering a pessimistic 50:50 background-signal operation, the halo DM and SNDM median longitude bands have an angular separation of roughly 35 degrees, which would allow for a discrimination between the two different models even in pessimistic background conditions.
The lower-mass models, on the other hand, present highly overlapping bands, with the 75th percentile SNDM line approaching the expected halo DM source direction at $\lambda = 0.5$, rendering the distinction of the two signals even more challenging.

While this is by no means a comprehensive study of the backgrounds, this analysis shows that a directional detector can still provide a positive DM discovery claim even in the presence of unmodelled backgrounds.

\section{Conclusions} 
\label{sec:conclusions} 

Using three sets of dark matter models, we compared signals that produce equivalent recoil spectra in a gas time-projection chamber despite their differing origin: two MeV-scale dark matter models originating from the galactic centre, boosted to mimic GeV-scale nuclear recoil signals, and non-relativistic WIMP nuclear recoils pointing back towards the constellation Cygnus. We generated recoil events from the directional recoil rates outlined in \ref{sec:dirn_dm}, and modelled detector effects using three performance benchmarks, namely ``ideal", ``realistic" and ``poor". These are characterised by an increasingly degrading angular resolution and head-tail effect recognition efficiency.

We quantified the number of events required to reject an isotropic background for each signal and detector performance. We found that isotropy can be rejected with fewer recoil events for the higher-mass models than for the lower-mass models, and that the halo DM case is the easiest to distinguish from isotropy in all cases: the higher-mass halo models require roughly four orders of magnitude fewer events than the galactic centre models, while the lower-mass models require roughly five orders of magnitude fewer. This follows from the directional rates from Fig.~\ref{fig:combined_dirn_rate}, where halo DM recoils are more strongly peaked toward the source direction.

We also quantified the number of events required to resolve the true source direction within 20 degrees by computing the median generated longitude across experiments for varying number events. The halo DM case again requires fewer events in both mass regimes, owing to its peaked recoil rate towards the source direction: assuming a realistic detector performance, the halo direction can be resolved with as few as $\sim$20 events for the higher-mass model and $\sim$60 events for the lower-mass model.

These results reflect a balance between intrinsic physical effects and instrumental ones. Higher-mass interactions produce higher-energy recoils, resulting in broader angular distributions on one hand, but better detector performances and, hence, measurements on the other. This latter effect dominates and leads to the higher-mass models outperforming the lower-mass ones overall, resulting in both the rejection of isotropy and the resolution of the true longitude with fewer events.

All of our analysis assumed an unrealistic background-free operation, so it is valuable to define a signal purity parameter $\lambda$ to estimate the effects of backgrounds in the detector. This is shown in Fig.~\ref{fig:sig_pur}, where the interquartile range of the median longitude for halo DM and SNDM is shown as a function of $\lambda$ for a realistic detector performance, assuming 50 signal events. Distinguishing the directions of the two signals becomes more difficult as the signal-to-background ratio decreases. In particular, resolving the two directions remains possible in the case of the higher-mass models, even at a pessimistic 50:50 background-to-signal operation, whereas the interquartile range bands for the lower-mass cases increasingly overlap, rendering their distinction more challenging.

Finally, we emphasise that this analysis is only possible because of directional information. The signals we examined were deliberately chosen to produce indistinguishable recoil energy spectra with identical event rates: a conventional detector would have no means of telling them apart. Directional detection offers the unique capability of identifying boosted MeV-scale dark matter signals that would otherwise be completely degenerate with GeV-scale non-relativistic WIMPs. This opens a window of detection capabilities that directionally insensitive experiments cannot access, enabling the investigation of the nature and properties of dark matter as well as its discovery. 

\begin{acknowledgments}

JLN, NFB \& ISA are supported by the Australian Research Council through the ARC Centre of Excellence for Dark Matter Particle Physics, CE200100008. ISA is supported by a University of Melbourne Graduate Research Scholarship. CL and CAJO
are supported by the Australian Research Council under the grant numbers DE220100225 and CE200100008.  

\end{acknowledgments}

\bibliographystyle{bibi.bst}
\bibliography{boosted.bib}

\end{document}